\documentclass[aps,prd,preprint,floatfix,nofootinbib]{revtex4}

\usepackage{graphicx,epsfig}

\newcommand{\GeV}{\unskip\,\mathrm{GeV}}

\newcommand{\lsim}
{\;\raisebox{-.3em}{$\stackrel{\displaystyle <}{\sim}$}\;}

\newcommand{\rmIm}{{\rm Im}}
\newcommand{\rme}{{\rm e}}
\newcommand{\rmcm}{{\rm cm}}
\newcommand{\lan}{{\langle}}
\newcommand{\ran}{{\rangle}}

\newcommand{\beq}{\begin{equation}}
\newcommand{\eeq}{\end{equation}}
\newcommand{\bea}{\begin{eqnarray}}
\newcommand{\eea}{\end{eqnarray}}
\newcommand{\nn}{\nonumber}
\newcommand{\ie}{{\it i.e.~}}

\begin{document}

\begin{flushright}
OITS-780\\
hep-ph/0606122\\
\end{flushright}

\bibliographystyle{revtex}

\title{Flavor Changing Effects in Family Nonuniversal $Z'$ Models}

\author{Cheng-Wei Chiang$^{1,2}$, N.G. Deshpande$^3$, J. Jiang$^3$}
\affiliation{$^1$Department of Physics, National Central University,
Chungli, Taiwan 320, R.O.C}
\affiliation{$^2$Institute of Physics, Academia Sinica, Taipei, Taiwan
115, R.O.C.}
\affiliation{$^3$Institute for Theoretical Science, University of Oregon,
Eugene, OR 97403, USA}\bigskip

\date{\today}

\begin{abstract}
  Flavor-changing and $CP$-violating interactions of $Z'$ to fermions
  are generally present in models with extra $U(1)$ gauge symmetry
  that are string-inspired or related to broken gauged family
  symmetry.  We study the consequences of such couplings in fermion
  electric dipole moments, muon $g-2$, and $K$ and $B$ meson mixings.
  From experimental limits or measured values, we constrain the
  off-diagonal $Z'$ couplings to fermions.  Some of these constraints
  are comparable or stronger than the existing constraints obtained
  from other observables.
\end{abstract}

\maketitle

\section{Introduction}
Additional heavy neutral $Z'$ gauge bosons have been extensively
studied in the literature.  They arise naturally in grand unified
models, superstring-inspired models~\cite{chaudhuri} and in models
with large extra dimensions~\cite{Masip:1999mk}.  There are stringent
limits on the mass of an extra $Z'$ from collider search experiments
at the Tevatron~\cite{Abe:1997fd}.  Precision data put limits on the
$Z$-$Z'$ mixing angle $\theta$~\cite{Erler:1999ub}.  Although these
limits are model-dependent, the typical constraints are $M_{Z'} >
O(500 \GeV)$ and $\theta < O(10^{-3})$.  Most studies have assumed
flavor universal $Z'$ couplings.  However, in intersecting D-brane
constructions, it is possible to have family nonuniversal $Z'$
couplings.  In extensions of the Standard Model (SM) with gauged
family symmetry, nonuniversal $Z'$ couplings also arise
naturally~\cite{Barger:1987hh}.  These couplings generally lead to
flavor-changing and $CP$-violating $Z'$ vertices, when quark and
lepton mixing is taken into account.  Flavor violating $Z$ couplings
can be induced through $Z$-$Z'$ mixing.  Experimental observables in
the flavor changing and $CP$ violating processes may be used to put
constraints on $Z'$ couplings.  One of the important searches that the
Large Hadron Collider (LHC) will undertake is to look for $Z'$ bosons.
It is therefore crucial to explore the parameter space of the allowed
couplings of such $Z'$ bosons.

In a recent paper~\cite{Langacker:2000ju}, Langacker and Pl{\"u}macher
have investigated the consequences of family nonuniversal $Z'$ gauge
boson.  They have considered several processes including $Z \to
\bar{q}_i q_j$, $l_i \to l_j l_k \bar{l}_m$, $\mu$-$e$ conversion,
radiative decays of $\mu \to e \gamma$ and $b\to s \gamma$, and meson
decays, etc.  In this paper we extend their analysis to include
constraints from electric dipole moments (EDMs) of electron and
neutron and muon $g-2$.  We re-analyze mass difference and $CP$
violation in $K$-$\bar{K}$ to emphasize the enhanced contributions
from left-right mixing terms.  For $B_d$ mixing, we find that it is
important to include an independent observable, that is not affected
by the $Z'$ effects, to improved the constraints.

In the Standard Model, the fermion EDMs are generated at three-loop or
higher order.  Thus the predicted value of less than $10^{-33} \rme
\cdot \rmcm$ is sevral orders of magnitude lower than the most
stringent bounds coming from electron and neutron EDM measurements.
However, in extensions of the SM, such as $Z'$ models with family
nonuniversal couplings, additional weak phases will allow fermion EDMs
to be generated at one-loop level, and thus they can be dominant.
Therefore, we can use EDM measurements to constrain $Z'$ flavor
changing couplings.

By exchanging a $Z'$ boson, oscillations of $K$, $B_d$ and $B_s$
mesons can occur via tree level diagrams, as compared to one-loop box
diagrams in the SM.  Some of the operators that occur in the $K$-$\bar
K$ mixing are enhanced so that very strong limits can be obtained from
$\Delta M_K$ and $\epsilon_K$ measurements.  In the $B_d$ system, the
limit on $|V_{ub}|$, the measurements of $\Delta M_{B_d}$ and
$\sin2\beta$ and the recent limit on $|V_{td}/V_{ts}|$ obtained from
$b \to d \gamma$ and $b \to s \gamma$~\cite{bellepreprint:2006-5} can
be combined to provide strong constraints on the $Z'$ flavor violating
couplings.  The recent measurements of $\Delta M_{B_s}$ at both D\O\
\cite{Abazov:2006dm} and CDF~\cite{Gomez-Ceballos:2006gj} have
generated much interest in flavor mixing in $B$ mesons
\cite{Blanke:2006ig}.  Several papers have studied $B_s$ mixing in the
context of $Z'$ models~\cite{Barger:2004qc, Cheung:2006tm,
Ball:2006xx, He:2006bk}.  Although the ratio $|\Delta M_{B_d} / \Delta
M_{B_s}|$ provides the best determination of $|V_{td}/V_{ts}|$ in the
SM, when new physics effects enter into both mass differences, the
ratio does not have an advantage over the individual mass differences
in constraining new physics variables.  We will discuss this point in
more details.

The paper is organized as the following, after the introduction, we
briefly describe our notations in Section~\ref{sec:form}.  We discuss
fermion EDMs in Section~\ref{sec:edm}, $K$-$\bar{K}$ mixing in
Section~\ref{sec:kk}, and $B$-$\bar{B}$ mixing in
Section~\ref{sec:bb}.  We conclude in Section~\ref{sec:con}.

\section{Formalism}
\label{sec:form}
We follow closely the formalism in Ref.~\cite{Langacker:2000ju}.  In
the gauge eigenstate basis, the neutral current Lagrangian can be
written as
\beq
{\mathcal L}_{\rm NC} = -eJ^{\mu}_{\mbox{\tiny em}} A_{\mu} - g_1
J^{(1)\,\mu} Z^{0}_{1,\mu} - g_2 J^{(2)\,\mu} Z^{0}_{2,\mu}\,,
\eeq
where $Z^0_1$ is the $SU(2)\times U(1)$ neutral gauge boson, $Z^0_2$
the new gauge boson associated with an additional Abelian gauge
symmetry and $g_1$ and $g_2$ are the corresponding gauge couplings.
The current associated with the additional $U(1)'$ gauge symmetry can
be generally written as
\beq
  J^{(2)}_{\mu} = \sum\limits_{i,j} \overline{\psi}_i \gamma_{\mu}
    \left[ \epsilon^{(2)}_{\psi_{L_{ij}}}P_L
     + \epsilon^{(2)}_{\psi_{R_{ij}}}P_R\right]\psi_j\,,
\eeq
where $\epsilon^{(2)}_{\psi_{{L,R}_{ij}}}$ is the chiral coupling of
$Z_2^0$ with fermions, and $i$ and $j$ run over quark flavors, and
similarly for leptons.  Flavor changing effects arise if
$\epsilon^{(2)}$ are nondiagonal matrices.  If the $Z_2^0$ couplings
are diagonal but family-nonuniversal, flavor changing couplings are
induced by fermion mixings.  When fermion Yukawa matrices $h_\psi$ are
diagonalized by unitary matrices $V^{\psi}_{R,L}$
\beq
  h_{\psi,{\rm diag}}=V^{\psi}_R\,h_{\psi}\,{V^{\psi}_L}^{\dagger}\,,
\eeq
the current associated with $Z_2^0$ is rewritten in the fermion mass
eigenstate basis
\beq
J^{(2)}_{\mu} = \sum_{i,j} \overline{\chi}_i \gamma_\mu \left[
B_{ij}^{\psi_L} P_L + B_{ij}^{\psi_R} P_R \right] \chi_j \,,
\eeq
with
\beq
  B^{\psi_L}_{ij}\equiv
    \left(V^{\psi}_L \epsilon^{(2)}_{\psi_L}
{V^{\psi}_L}^{\dagger}\right)_{ij}\,,
    \qquad\mbox{and}\qquad
  B^{\psi_R}_{ij}\equiv
    \left(V^{\psi}_R \epsilon^{(2)}_{\psi_R}
{V^{\psi}_R}^{\dagger}\right)_{ij}\,.
  \label{Bij}
\eeq
In the following sections we simplify $B_{ij}^{\psi_{L,R}}$ as
$B_{ff'}^{L,R}$, with $f$ and $f'$ specifying the flavors of quarks
and leptons explicitly and the $L,R$ superscripts indicating
left-handed or right-handed couplings.  For example, $B_{13}^{d_L}$
will be written as $B_{db}^L$.  In general, $B_{ff'}^L$ and
$B_{ff'}^R$ can be complex and have independent phases.

\section{Electric Dipole Moments}
\label{sec:edm}

\begin{figure}[htb]
\centering
\includegraphics[width=6.7cm]{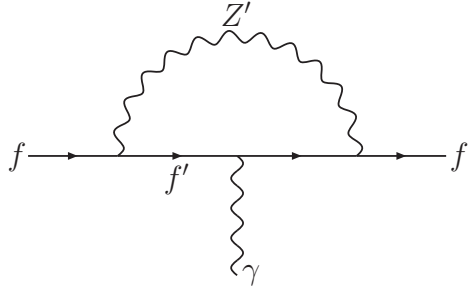}
\caption{Feynman diagram of fermion EDMs at one-loop level through
  flavor changing complex $Z'ff'$ coupling.}
\label{fig:feyndiag}
\end{figure}

In the SM, for a dipole operator, the weak phase exactly cancels out
in the one-loop diagrams.  Hence there is no contribution to fermion
EDMs at one-loop level.  In comparison, there can be independent
phases involving the left-handed and the right-handed $Z'$ couplings.
These complex phases can contribute to fermion EDMs through the
one-loop diagram shown in Fig.~\ref{fig:feyndiag}, where $f$ and $f'$
indicate fermions of different flavors.

For fermion $f$, the contribution to its EDM from the
Fig.~\ref{fig:feyndiag} is evaluated to be,
\bea
d_f = -
\frac{1}{16 \pi^2} g_2^2 Q_{f} e \frac{m_{f'}}{m_{Z'}^2} {\rm
Im}({B_{f f'}^{L}}^* B_{f f'}^R) \int_0^1 dx \frac{a x^4 + 4
x(1-x)}{a x^2 + b x + 1}\,,
\eea
where, $Q_f$ is the charge and $m_f$ is the mass of the external
fermion, $m_{f'}$ is the mass of the internal fermion, $m_{Z'}$ is
$Z'$ boson mass, and $a = m_f^2/m_{Z'}^2$ and
$b=(m_{f'}^2-m_f^2)/m_{Z'}^2-1$.  Note the contribution is non-zero
only if both $B_{f f'}^L$ and $B_{f f'}^R$ are non-zero, at least one
of them is complex, and their phases do not cancel.  In the
approximation of external quark mass being much less than the $Z'$
mass,
\ie, $m_f \ll m_{Z'}$, the above equation can be simplified to be
\beq
d_f = -
\frac{1}{8 \pi^2} g_2^2 Q_f e \frac{m_{f'}}{m_{Z'}^2} {\rm
Im}({B_{f f'}^{L}}^* B_{f f'}^R) \frac{1 -c^2 + 2 c \log c}{(1-c)^3}\,,
\eeq
in which, $c = m_{f'}^2/m_{Z'}^2$.  If the internal quark mass is
again much less than $m_{Z'}$, \ie, $m_{f'}\ll m_{Z'}$, then the
equation can be further simplified into
\bea
d_f &=& - \frac{1}{8 \pi^2} g_2^2 Q_f e \frac{m_{f'}}{m_{Z'}^2} {\rm
Im}({B_{f f'}^{L}}^* B_{f f'}^R) \nn \\
&=& - \frac{g_1^2}{8 \pi^2}
Q_f e \frac{m_{f'}}{m_Z^2}\,y\,{\rm Im}({B_{f f'}^{L}}^* B_{f f'}^R)\,.
\eea
where parameter $y$ is defined as
\beq
y \equiv \left(\frac{g_2}{g_1}\right)^2 \frac{M_Z^2}{M_{Z^\prime}^2}\,.
\eeq
Now we can apply this result to electron and $u$ and $d$ quark EDMs.
For electrons, both the diagrams with internal $\mu$ and internal
$\tau$ contribute.  By requiring both contributions to be less than
electron EDM constraint $d_e < 1.4 \times 10^{-27} {\rm e} \cdot {\rm
cm}$~\cite{Regan:2002ta}, we get
\beq
\label{eq:l12}
y\, \rmIm({B_{e\mu}^{L}}^* B_{e\mu}^{R}) < 1 \times 10^{-6}\,,
\eeq
\beq
\label{eq:l13}
y\, \rmIm({B_{e\tau}^{L}}^* B_{e\tau}^{R}) < 7 \times 10^{-8}\,.
\eeq
The constraint on $B_{e\tau}$ is stronger simply because the
contributions to the EDMs are proportional to the internal fermion
masses.

The strongest bounds on $B_{e\mu}^{L}$ and $B_{e\mu}^{R}$ come from
the non-observation of coherent $\mu$-$e$
conversion~\cite{Langacker:2000ju} by the Sindrum-II
Collaboration~\cite{Wintz:1998rp}, as the small mixing between $Z$ and
$Z'$ can induce such conversion process,
\beq
w^2 (|B_{e\mu}^{L}|^2 + |B_{e\mu}^{R}|^2) < 4 \times 10^{-14}\,,
\label{eq:pll12}
\eeq
where $w = g_2/g_1 \sin\theta \cos\theta (1 - m_Z^2/m_{Z'}^2)$ and
$\theta$ is the $Z$-$Z'$ mixing angle.  In the most interesting case
of a TeV-scale $Z'$ with small mixing, $\theta \lsim 10^{-3}$, $y$ and
$w$ are of the same order, and $y \approx w \approx 10^{-3}$.  This is
the case we assume in comparing the constraints from different
processes.  It is difficult to directly compare the constraints in
Eq.~(\ref{eq:l12}) and Eq.~(\ref{eq:pll12}), since the former depends
on the phase difference between $B_{e\mu}^{L}$ and $B_{e\mu}^{R}$ and
the later on the absolute values.  As $|B_{e\mu}^{L,R}|$ become as
small as in Eq.~(\ref{eq:pll12}), the constraint in Eq.~(\ref{eq:l12})
becomes unimportant.  In this sense, we say the coherent $\mu$-$e$
conversion provides a stronger constraint on $B_{e\mu}^{L,R}$ than the
electron EDM.  The decay $\tau \to 3 e$ provides the best constraint
on flavor violating $Z'e\tau$ coupling
\beq
w^2 ( |B_{e\tau}^{L}|^2 + |B_{e\tau}^{R}|^2) < 2 \times 10^{-5}\,.
\eeq
In this case, the constraint from electron EDM, Eq.~(\ref{eq:l13}), is
more stringent, although it depends on phases.

We can also apply the same constraint on quark EDMs inferred from the
neutron EDM.  Barring possible cancellations, we require each diagram
contributes less than the experimental limit $d_n < 3.0 \times
10^{-26}
\rme \cdot \rmcm$~\cite{Baker:2006ts}, we get the following constraints
\bea
y\, \rmIm({B_{uc}^{L}}^* B_{uc}^{R}) &<& 3 \times 10^{-6} \\ y\,,
\rmIm({B_{ds}^{L}}^* B_{ds}^{R}) &<& 5 \times 10^{-5}\,,
\label{eq:ud12}\\
y\, \rmIm({B_{ut}^{L}}^* B_{ut}^{R}) &<& 2  \times 10^{-8}\,, \\
y\, \rmIm({B_{db}^{L}}^* B_{db}^{R}) &<& 2  \times 10^{-6}\,.
\label{eq:ud13}
\eea
Eq.~(\ref{eq:ud12}) gives a weaker bound on $B_{ds}^{L,R}$ than those
from $K_L \to \mu^+\mu^-$~~\cite{Eidelman:2004wy} and $K_L \to \pi^0
\mu^+ \mu^-$~\cite{Whitmore:1999vt} decays
\bea
&& w^2 | {\rm Re}B_{ds}^{R}-{\rm Re}B_{ds}^{L}|^2 < 3 \times 10^{-11}\,,
\nn \\
&& w^2 | {\rm Im}B_{ds}^{R}+{\rm Im}B_{ds}^{L}|^2 < 5 \times 10^{-11}\,.
\eea
At the same time, the constraint in Eq.~(\ref{eq:ud13}) are relevant,
compared to bounds that come from $B^0$ decay into a $\mu^+\mu^-$ pair
\cite{D0FPCP},
\beq
w^2 |B_{db}^{L,R}|^2 < 10^{-5}\,.
\eeq

The same diagram in Fig.~\ref{fig:feyndiag}, with the external fermion
being $\mu$, will contribute to muon $g-2$.  The general expression
for the contributions from the $Z'$ diagram is given in
Ref.~\cite{Lynch:2001zr}.  As the external and internal leptons masses
are far smaller than the $Z'$ mass, the dominant contribution becomes
\beq
a_\mu^{Z'} = - \frac{y}{4 \pi^2} \frac{g_1^2}{m_Z^2} m_\mu m_\tau {\rm
Re}({B_{\mu\tau}^L}^* B_{\mu\tau}^R)\,.
\eeq
If we demand this contribution to be less the the difference between
the experimentally measured value and the Standard Model
prediction~\cite{Czarnecki:2005th}, $\Delta a_\mu < 250 \times
10^{-11}$, we get
\beq
y \,{\rm Re}({B_{\mu\tau}^{L}}^* B_{\mu\tau}^{R}) < 1 \times
10^{-2}\,.
\eeq
In the aforementioned small mixing and TeV-scale $Z'$ case, this
constraint on $B_{\mu\tau}$ is as strong as the one derived from $\tau
\to 3 \mu$ decay~\cite{Langacker:2000ju},
\beq
w^2 (|B_{\mu\tau}^{L}|^2 + |B_{\mu\tau}^{R}|^2) < 10^{-5}\,.
\eeq
The contribution from the diagram with electron in the loop are
suppressed by the much lighter electron mass, thus it does not provide
a useful constraint.

\section{$K$-$\overline{K}$ Mixing}
\label{sec:kk}

The off-diagonal element $M_{12}$ in the neutral $K$-$\overline{K}$
mixing mass matrix is related to the $|\Delta S| = 2$ effective
Hamiltonian by
\beq
2 m_K M_{12}^*
= \lan \overline{K}^0 | {\cal H}_{\rm eff}^{|\Delta S| = 2} | K^0 \ran ~.
\eeq
With the definition
\beq
\lan \overline{K}^0 | [\bar s \gamma_{\mu} (1 - \gamma_5) d]
    [\bar s \gamma^{\mu} (1 - \gamma_5) d] |
K^0 \ran \equiv
\frac83 B_K f_K^2 m_K^2 ~,
\eeq
one obtains within the SM~\cite{Buchalla:1995vs}
\bea
M_{12}^{\rm SM} &=&
\frac{G_F^2}{16\pi^2} M_W^2
\left[ (V_{cd}^*V_{cs})^2 \eta_1 S_0(x_c) + (V_{td}^*V_{ts})^2 \eta_2 S_0(x_t)
  + 2 (V_{cd}^*V_{cs}) (V_{td}^*V_{ts}) \eta_3 S_0(x_c,x_t)
\right]
\nn \\
&& \quad \times \frac43 B_K^{LL} f_K^2 m_K ~,
\eea
where the QCD factors $\eta_1 \simeq 1.38$, $\eta_2 \simeq 0.57$, and
$\eta_3 \simeq 0.47$, and the Inami-Lim functions~\cite{Inami:1980fz}
$S_0(x)$ and $S_0(x,y)$ can be found, for example, in
Ref.~\cite{Buchalla:1995vs}.  The renormalization scale and scheme
invariant bag parameter is
\beq
B_K^{LL} = \alpha_s^{(4)}(\mu)^{-2/9} \left[ 1 + 1.895
  \frac{\alpha_s^{(4)}(\mu)}{4\pi}\right] B_K^{LL}(\mu)\,,
\eeq
with the same factor for left-right mixing bag parameters $B_{K1}^{LR}$ and
$B_{K2}^{LR}$.  We will take the following numerical values: $B_K^{LL}(2~{\rm
  GeV}) = 0.69 \pm 0.21$, $B_{K1}^{LR}(2~{\rm GeV}) = 1.03 \pm 0.06$, and
$B_{K2}^{LR}(2~{\rm GeV}) = 0.73 \pm 0.10$~\cite{Allton:1998sm}, $f_K = 159.8
\pm 1.5$ MeV, and $m_K = 497.648 \pm 0.022$ MeV~\cite{Eidelman:2004wy}.  The
short-distance contribution to the mass difference between the two
mass eigenstates of kaons, $\Delta M_K$, is
\beq
\Delta M_K^{\rm SD} = 2 {\rm Re} M_{12} ~.
\eeq
The $CP$ violation is measured by the parameter
\beq
\epsilon_K = \frac{e^{i \pi / 4}}{\sqrt{2} \Delta M_K} {\rm Im} M_{12} ~.
\eeq
Their experimental values are given in~\cite{Eidelman:2004wy} as
$\Delta M_K = 0.5292 \times 10^{10}~{\rm s}^{-1}$ and $|\epsilon_K| =
2.284 \times 10^{-3}$.

The general set of $|\Delta S| = 2$ operators relevant for our
discussions is:
\begin{eqnarray}
  && O^{LL}
  = [\bar s \gamma_{\mu} (1 - \gamma_5) d]
    [\bar s \gamma^{\mu} (1 - \gamma_5) d] ~, \nn \\
  && O_1^{LR}
  = [\bar s \gamma_{\mu} (1 - \gamma_5) d]
    [\bar s \gamma^{\mu} (1 + \gamma_5) d] ~, \nn \\
  && O_2^{LR}
  = [\bar s (1 - \gamma_5) d] [\bar s (1 + \gamma_5) d] ~, \nn \\
  \label{eq:ops}
  && O^{RR}
  = [\bar s \gamma_{\mu} (1 + \gamma_5) d]
    [\bar s \gamma^{\mu} (1 + \gamma_5) d] ~.
\end{eqnarray}
As seen previously, only the operator $O^{LL}$ contributes to
$K$-$\overline{K}$ mixing in the SM due to its chiral structure.  The
other three operators appear in the $Z'$ models because the left- and
right-handed couplings and operators mix through renormalization group
(RG) evolution.  The RG running of the Wilson coefficients $C(\mu)$
from the $M_W$ scale down to the lattice scale $\mu_L$, where we will
match with the lattice results of the associated hadronic matrix
elements, can be schematically written as
\bea
C(\mu_L) = U(\mu_L,M_W) C(M_W) ~.
\eea
Details of computing the evolution matrix $U(\mu_L,M_W)$ are given in
Ref.~\cite{Ciuchini:1997bw}.  Here we only provide the numerical
values of the relevant evolution matrices:
\bea
U_{LL}^K &=& U_{RR}^K
\simeq 0.788 ~, \\
U_{LR}^K &=& U_{RL}^K
\simeq \left(\begin{array}{cc}
    0.906 & -0.087 \\ -1.531 & 3.203
  \end{array}\right) ~.
\eea
In determining these results, we have used only the central value of
$\alpha_s(M_Z) = 0.118$ for a 5-quark effective theory and chose the
lattice scale $\mu_L = 2$ GeV.  The additional contribution from the
$Z'$ model with only the left-handed couplings is
\bea
M_{12}^{LL} =
\frac{G_F}{\sqrt{2}} y U_{LL}^K \left(B_{ds}^{L} \right)^2
\frac43 B_K^{LL} f_K^2 m_K \,,
\label{eq:kll}
\eea
and the expression that also includes the right-handed coupling is
\bea
M_{12}^{LR} = &&
\frac{G_F}{\sqrt{2}}\, y f_K^2 m_K \left\{ \frac43 U_{LL}^K
\left(B_{ds}^{L}\right)^2 B_K^{LL} + \frac43 U_{RR}^K
\left( B_{ds}^{R}\right)^2 B_K^{RR} \right. \nn \\  
&& \left. + \left(\frac{m_K}{m_s + m_d}\right)^2 B_{ds}^L
B_{ds}^{R} \left[ -\frac{4}{3} U_{LR}^K(1,1) B_{K1}^{LR} + 2
U_{LR}^K(2,1) B_{K2}^{LR} \right] \right\} \,.
\label{eq:klr}
\eea
If we constrain the contribution to $\Delta M_K$ from $M_{12}^{LL}$ to
be less than the currently measure experimental value, we get the
bound,
\beq
y | \,{\rm Re}\left( B_{ds}^{L}\right)^2|< 2 \times 10^{-8}\,.
\eeq
After including the $RR$ and the $LR$ mixing terms, the constraint
becomes
\beq
y | 0.01 \, {\rm Re}\left[\left( B_{ds}^{L}\right)^2 + \left(
B_{ds}^{R}  \right)^2 \right] - {\rm Re}\left(B_{ds}^{L}
B_{ds}^{R}\right) | < 2 \times 10^{-10}\,.
\eeq
Keeping the dominant term, we can simplify the equation into
\beq
y | {\rm Re}\left(B_{ds}^{L} B_{ds}^{R}\right) | < 2 \times
10^{-10}\,.
\eeq

The theoretical uncertainty on $\epsilon_K$ within the Standard Model
is mainly due to the uncertainty of the bag parameter $B_K$ and it is
estimated to be about $30\%$~\cite{Allton:1998sm}.  If we require the
contribution from $Z'$ is less than the theoretical error associated
with the SM prediction, we have
\bea
\frac{|\epsilon_K^{Z'}|}{|\epsilon_K^{exp}|} &=& \frac{|{\rm Im}
M_{12}^{Z'}|}{\sqrt{2}\Delta M_K^{\rm exp} \epsilon_K^{\rm exp}} \nn
\\ 
&\approx& 1 \times 10^{12} \,y\, | 0.01\, {\rm Im}\left[\left(
B_{ds}^{L} \right)^2 + \left( B_{ds}^{R}\right)^2 \right] - {\rm
Im}\left( B_{ds}^{L} B_{ds}^{R}\right) | < 0.3\,.
\eea
Assuming only the $LL$ coupling exists, the constraint becomes
\beq
y \,{\rm Im}\left( B_{ds}^{L}\right)^2 < 3 \times
10^{-11}\,.
\eeq
When both left-handed and right-handed couplings contribute, we can
ignore the $LL$ and $RR$ terms, and obtain
\beq
y \,{\rm Im}\left( {B_{ds}^{L}} B_{ds}^{R}\right) < 3 \times
10^{-13}\,.
\eeq
In comparison with Ref.~\cite{He:2004it}, the stronger bound here is
due to the chiral and renormalization enhancement in the left-right
mixing terms.

\section{$B_d$-$\overline{B}_d$ Mixing}
\label{sec:bb}

Similar to $K$-$\overline{K}$ mixing, chiral couplings of the $Z'$
with the $b$ and $d$ quarks can induce $B_d$-$\overline{B}_d$ mixing.
In the SM, the off-diagonal element in the $B_d$ meson mass matrix is
given by~\cite{Buchalla:1995vs}
\beq
M_{12}^{\rm SM} = \frac{G_F^2}{16\pi^2} M_W^2 (V_{tb}^*V_{td})^2
\eta_B  \frac{4}{3} B_{B}^{LL} f_{B}^2 m_{B} S_0(x_t)\,,
\eeq
where the QCD factor $\eta_B \simeq 0.55$, $S_0(x_t) =
2.463$~\cite{Buchalla:1995vs}, and $m_{B} = 5.2794 \pm 0.0005
\GeV$~\cite{Eidelman:2004wy}.  The renormalization scale invariant bag
parameter is
\beq
B_{B}^{LL} = \alpha_s^{(5)}(\mu)^{-6/23} \left[ 1 + 1.627
  \frac{\alpha_s^{(5)}(\mu)}{4\pi}\right] B_{B}^{LL}(\mu) ~,
\eeq
with similar expressions for $B_{B1}^{LR}$ and $B_{B2}^{LR}$, bag
parameters for left-right mixing operators.  The bag parameters in the
$\overline{\rm MS}$ scheme are evaluated on the lattice with quenched
approximation~\cite{Becirevic:2001xt} and they are:
$B_{B}^{LL}(4.6~\GeV) = 0.87 \pm 0.06$, $B_{B1}^{LR}(4.6~{\rm GeV}) =
1.72 \pm 0.12$, and $B_{B2}^{LR}(4.6~{\rm GeV}) = 1.15 \pm 0.6$.  The
decay constant is $f_{B} = 173 \pm 23~{\rm MeV}$.

It is a common practice to determine $\sin2\beta$ from the
time-dependent $CP$ asymmetry of the $b \to c \bar c s$ processes
because the decay amplitudes are dominated by color-suppressed
tree-level processes and thought to be least affected by new physics
contributions~\cite{Hou:2006du}.  Within the Standard Model,
$\sin2\beta$ is related to the CKM matrix elements
\beq
\beta = \arg\left(-\frac{V_{cd} V_{cb}^*}{V_{td} V_{tb}^*}\right)~.
\eeq
Both $\Delta M_B$ and $\sin2\beta$, determined from all charmonium
modes, are measured at Belle~\cite{Abe:2004mz} and
BaBar~\cite{Aubert:2004zt} and the world average~\cite{:2006bi} are
\bea
\Delta M_B &=& 0.507 \pm 0.005\,{\rm ps}^{-1}\,,\\
\sin2\beta &=& 0.687 \pm 0.032 ~.
\eea

A set of $|\Delta B| = 2$ operators can be obtained by simply
replacing $\bar s$ with $\bar b$ in Eq.~(\ref{eq:ops}).  Following
Ref.~\cite{Ciuchini:1997bw}, we calculate the evolution matrices,
\bea
U_{LL}^B &=& U_{RR}^B \simeq 0.842\,, \\
U_{LR}^B &=& U_{RL}^B
\simeq \left(\begin{array}{cc}
    0.921 & -0.041 \\ -0.882 & 2.081
  \end{array}\right) ~.
\eea
The contributions from $Z^\prime$ with purely left-handed coupling and
with both left-handed and right-handed couplings to the off-diagonal
$M_{12}^B$ are similar to Eq.~(\ref{eq:kll}) and Eq.~(\ref{eq:klr})
with simple replacements of parameters
\bea
M_{12}^{LL} &=&
\frac{G_F}{\sqrt{2}} y U_{LL}^B \left(B_{db}^{L} \right)^2
\frac43 B_B^{LL} f_B^2 m_B \,, \\
M_{12}^{LR} &=&
\frac{G_F}{\sqrt{2}}\, y f_B^2 m_B \left\{ \frac43 U_{LL}^B
\left(B_{db}^{L}\right)^2 B_B^{LL} + \frac43 U_{RR}^B
\left( B_{db}^{R}\right)^2 B_B^{RR} \right. \nn \\  
&& \left. + \left(\frac{m_B}{m_d + m_b}\right)^2 (B_{db}^{L}
B_{db}^{R}) \left[ - \frac{4}{3}
U_{LR}^B(1,1)  B_{B1}^{LR} +  2 U_{LR}^B(2,1) B_{B2}^{LR}\right] \right\}\,.
\eea

In the presence of $Z^\prime$ contributions, the weak phase thus measured
should be an effective one, with
\beq
\beta_{\rm eff} = - \frac12 \arg (M_{12}^{\rm SM} + M_{12}^{Z'} )~.
\eeq
The measured $\Delta M_{B}$ and $\sin2\beta_{\rm eff}$ may both
contain contributions from $Z^\prime$.  Therefore, assuming existence
of new physics, they cannot be used to determine the SM $V_{td}$.
Without the accurate determination from $B_d$ mixing and decay,
information on the Wolfenstein parameters $\rho$ and
$\eta$~\cite{Wolfenstein:1983yz} can only be derived from two sources.
On the one hand, we can deduce constraints on $\sqrt{\rho^2 + \eta^2}$
from $|V_{ub}| = (4.05 \pm 0.52 ) \times 10^{-3}$, $|V_{cb}| = (41.4
\pm 2.1 ) \times 10^{-3}$ and $|V_{cd}| = 0.224 \pm
0.014$~\cite{Charles:2004jd} and allow $\rho$ and $\eta$ values to
vary within this constraint.  $|V_{ub}|$ and $|V_{cb}|$ are determined
from semileptonic decays of $B$ mesons.  $|V_{cd}|$ can be deduced
from neutrino and antineutrino production of charm off valence $d$
quarks.  From $|V_{cd}|$, $|V_{cb}|$ and $|V_{ub}|$ we have
\beq
\left| \frac{V_{ub}}{V_{cd} V_{cb}} \right| = |\rho - i \eta | =
0.437\pm 0.066\,.
\label{eq:con1}
\eeq
On the other hand, the ratio $|V_{td}/V_{ts}|$ has recently been
determined at BELLE~\cite{bellepreprint:2006-5} through $b \to d
\gamma$ decays.  Its value is found to be within the interval of
$0.142 < |V_{td}/V_{ts}| < 0.259$ at a $95\%$ confidence level.  More
importantly, diagrams involving $Z'$ that contribute to the $b \to d
\gamma$ or $b\to s \gamma$ process are not only loop suppressed but
also mass suppressed.  Therefore, the bound on $|V_{td}/V_{ts}|$
provides an additional constraint on the SM $\rho$ and $\eta$
parameter space.  The constraint derived from $|V_{td}/V_{ts}|$,
combined with $|V_{cd}|$, gives the ratio and its $1\sigma$ range
\beq
\left|\frac{V_{td}}{V_{cd} V_{ts}}\right| = | 1- \rho -i \eta | =
0.888 \pm 0.163\,.  
\label{eq:con2}
\eeq
Note we have used the $95\%$ confidence level bound to derive the
$1\sigma$ error and turned asymmetric errors to symmetric ones
assuming they are Gaussian.

In $\Delta M_d^{\rm exp}/\Delta M_s^{\rm exp}$, the ratio of the
hadronic parameters $(f_{B_d}^2 B_{B_d})/(f_{B_s}^2 B_{B_s})$ is more
accurately known than individual hadronic parameters.  It may seem
that the ratio would provide a better determination of the related
$Z'$ couplings.  This is not so when $Z'$ effects enter both $\Delta
M_d$ and $\Delta M_s$.  While trying to constrain $B_{db}$ from the
ratio, we need to know $B_{sb}$.  The hadronic uncertainties re-enter
in the form of uncertainty on $B_{sb}$~\cite{Cheung:2006tm}.  Hence,
we use $\Delta M_{db}$ and $\sin2\beta_{eff}$ to constrain $B_{db}$,
together with the bounds on $\rho$ and $\eta$ discussed above.  Note
that, in the $B_s$ system, the $CP$-violating parameter $\sin2\phi_s$
can be combined with $\Delta M_{B_s}$ to determine new physics
parameters~\cite{Barger:2004qc,Ball:2006xx}.

To be specific, we consider only the $LL$ couplings in the following
discussions.  We now rewrite $M_{12}$ in simple forms of $\rho$,
$\eta$ and $B_{db}$,
\bea
M_{12}^{\rm SM}(B_d)
&=& \frac{G_F^2}{12\pi^2} M_W^2 \eta_B
B_{B_d}^{LL} f_{B_d}^2 m_{B_d} S_0(x_t) (V_{tb}^*V_{td})^2 \nn \\
&=& C_1 [1-(\rho + i \eta)]^2 \,,\\
M_{12}^{LL}
&=&
\frac{G_F}{\sqrt{2}} y U_{LL}^B \left(B_{db}^{L} \right)^2
\frac43 B_{B_d}^{LL} f_{B_d}^2 m_{B_d} \nn\\
&=& C_2 \,y \left(B_{db}^{L} \right)^2\,,
\eea
with
\bea
&& C_1 \equiv \frac{G_F^2}{12\pi^2} M_W^2 \eta_B B_{B_d}^{LL} f_{B_d}^2
m_{B_d} S_0(x_t) A^2 \lambda^6 = (1.823 \pm 0.52) \times 10^{-13}~{\rm
GeV}\,, \\
&& C_2 \equiv \frac43 \frac{G_F}{\sqrt{2}} U_{LL}^B B_{B_d}^{LL}
f_{B_d}^2 m_{B_d} = (1.947 \pm 0.53) \times 10^{-6} \GeV \,,
\eea
and the ratio $C_2/C_1 = 1.068 \pm 0.085 \times 10^7$.  Here we take
$A = 0.801 \pm 0.024$ and $\lambda = 0.2262 \pm
0.0020$~\cite{Charles:2004jd}.  The observed $\Delta M_B$ and
$\sin2\beta$ render
\bea
&& 2 C_1 \left|[1-(\rho + i \eta)]^2 +
\frac{C_2}{C_1} \,y
\left(B_{db}^{L} \right)^2\right| = (3.337 \pm 0.033)\times 10^{-13}
\GeV \,,
\label{eq:con3}
\eea
\vspace*{-1cm}
\bea
&& -\arg\{[1-(\rho + i \eta)]^2 +
\frac{C_2}{C_1} \,y
\left(B_{db}^{L} \right)^2\} = 0.757_{-0.043}^{+0.045}\,.
\label{eq:con4}
\eea
We try to get limits on $y (B_{db})^2$ based on the four
conditions in Eq.(\ref{eq:con1}), (\ref{eq:con2}), (\ref{eq:con3}) and
(\ref{eq:con4}).

\begin{figure}[htb]
\centering
\includegraphics[width=8cm]{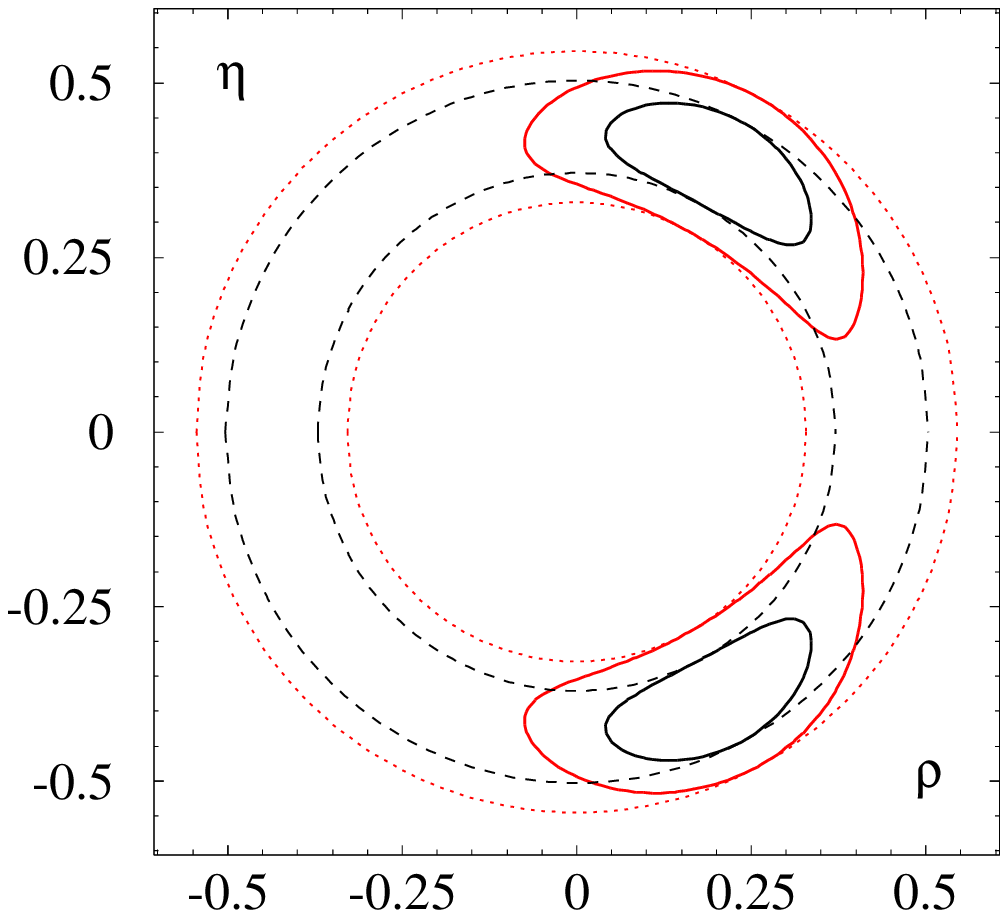}
\includegraphics[width=8cm]{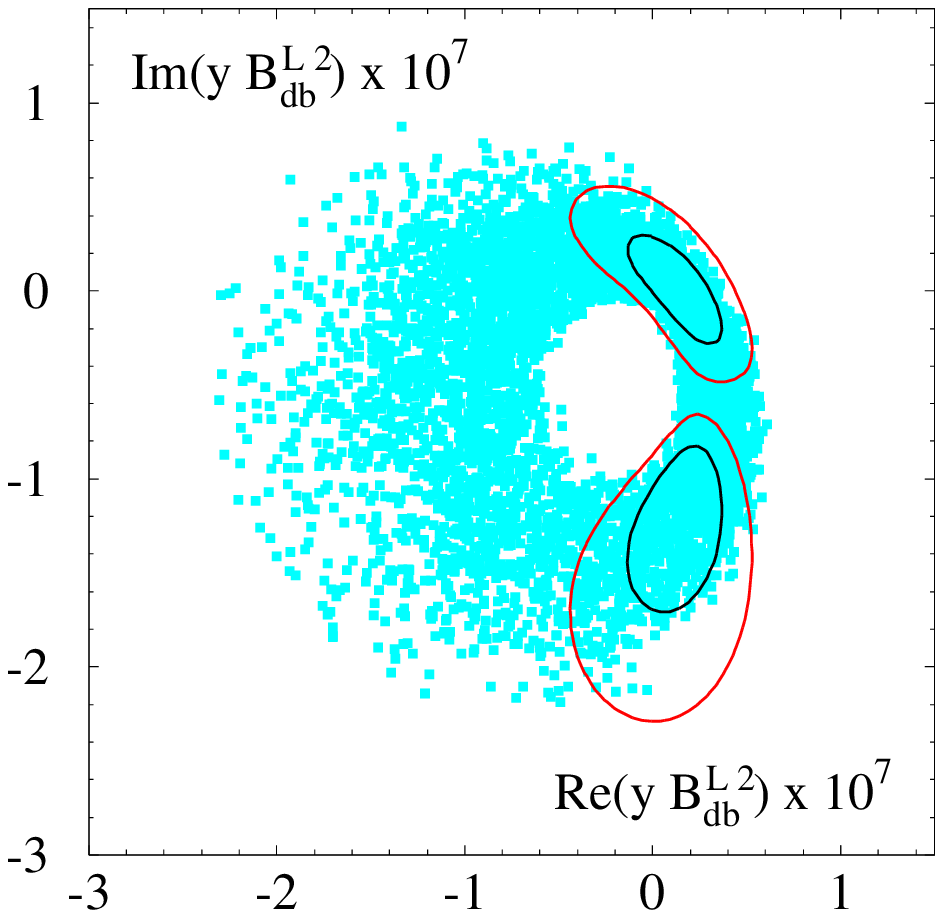}
\caption{The allowed $\rho$ and $\eta$ values (left) when both SM and $Z'$
  contribute to $B_d$ mixing, and the allowed ranges for $y {\rm
  Im}({B_{db}^{L}})^2$ and $y {\rm Re}({B_{db}^{L}})^2$ (right).  For
  the dashed ($1 \sigma$) and dotted ($1.64 \sigma$) contours on the
  left and the scattered points ($1\sigma$) on the right, the
  constraint on $|V_{td}/V_{ts}|$ is not imposed.  For the solid
  contours, black for $1\sigma$ and red for $1.64 \sigma$, the
  constraint on $|V_{td}/V_{ts}|$ is imposed.}
\label{fig:zprimell}
\end{figure}

In Fig.~\ref{fig:zprimell}, we show the allowed ranges in the
$\rho$-$\eta$ plane and in the plane of the $Z'$ parameters $y {\rm
Re}({B_{db}^{L}})^2$ and $y {\rm Im}({B_{db}^{L}})^2$.  We show the
results before imposing the $|V_{td}/V_{ts}|$ constraint, \ie,
Eq. (\ref{eq:con2}), with the dashed ($1\sigma$) and dotted
($1.64\sigma$) contours in the left plot, and with the scattered
points in the right plot.  Because of the additional $Z'$
contributions, $\rho$ and $\eta$ are allowed to take all possible
values allowed by the $|V_{ub}|$, $|V_{cd}|$ and $|V_{cb}|$
measurements, and the corresponding allowed range for $Z'$ parameters
are approximately $-2 \times 10^{-7} < y {\rm Re}({B_{db}^{L}})^2 < 1
\times 10^{-7}$ and $-2 \times 10^{-7} < y {\rm Im}({B_{db}^{L}})^2 <
1 \times 10^{-7}$.  However, after imposing the $|V_{td}/V_{ts}|$
constraint, the allowed region of $\rho$ and $\eta$ as well as those
of $ y {\rm Re}({B_{db}^{L}})^2$ and $y {\rm Im}({B_{db}^{L}})^2$
improve significantly, as shown by the solid black ($1\sigma$) and red
($1.64\sigma$) contours in both plots of Fig.~\ref{fig:zprimell}.
Just from the four conditions listed above, the $\eta < 0$ region is
allowed, leaving a twofold ambiguity on the allow regions.  Under the
assumption that $Z'$ is not the dominant contributions in
$\epsilon_K$, we can use the $\epsilon_K$ measurement to exclude the
$\eta < 0$ region.  From another point of view, for the lower regions
in both plots, the large $Z'$ contributions have to be canceled by the
SM contributions to reproduce $\Delta M_B$ and $\sin2\beta$
measurements.  The lower regions are thus less natural, and we limit
ourself to the upper regions.  Hence, when only left-handed couplings
are present, the bounds can be estimated from the right plot of
Fig.~\ref{fig:zprimell} to be
\bea
&&  y|{\rm Re}(B_{db}^{L})^2| < 5 \times 10^{-8}\,, \\
&&  y| {\rm Im}(B_{sd}^{L})^2| < 5 \times 10^{-8}~.
\eea
When both left-handed and right handed couplings are included, the
constraints are
\bea
&&  y| {\rm Re}[(B_{db}^{L})^2 + (B_{db}^{R})^2]
- 3.8 {\rm Re} ( B_{db}^{L}B_{db}^{R})| <  5 \times 10^{-8}\,, \\
&&  y| {\rm Im}[(B_{db}^{L})^2 + (B_{db}^{R})^2]
- 3.8 {\rm Im} ( B_{db}^{L}B_{db}^{R})| <  5 \times 10^{-8}\,,
\eea
which are less illuminating because of the possible cancellation among
different terms.

\section{conclusions}
\label{sec:con}

Flavor changing and $CP$ violating processes are natural consequences
of family-nonuniversal $Z'$ models, and they can manifest in
observables such as EDMs, muon $g-2$ and meson mixings.  We have
studied constraints on $Z'$ couplings from electron and neutron EDMs,
muon $g-2$, $K$ and $B$ meson mixing and $CP$ violation.

We presented the general expression for the fermion EDM generated by a
one-loop diagram induced by the $Z'$ boson.  In the approximation that
both internal and external fermion masses are much smaller than the
$Z'$ mass, the EDM is a simple quantity proportional to $Z'$ couplings
and the internal fermion mass.  We obtained the constraints on the
chiral couplings to $Z'$ by requiring each individual contribution to
be within the experimental limits of electron and neutron EDMs.
Derived From the electron EDM, the constraint on $B_{e\mu}^{L,R}$ is
weaker than that from the $\mu$-e conversion, while the constraint on
$B_{e\tau}^{L,R}$ is stronger than that from the $\tau \to 3 e$ decay.
From the neutron EDM, bounds on $B_{ds}^{L,R}$ are not as strong as
those imposed by the $K_L \to \mu^+\mu^-$ and $K_L \to \pi^0 \mu^+
\mu^-$ decays.  However, bounds on $B_{db}^{L,R}$ are stronger than
bounds from the $B^0$ to $\mu^+ \mu^-$ decay.  Because the EDMs are
proportional to the internal fermion masses, they provide better
constraints on couplings involving heavier leptons and quarks.
Requiring the $Z'$ contribution to muon $g-2$ to be less than the
discrepancy between theoretical and experimental values, we obtained
comparable limits on $B_{\mu\tau}^{L,R}$ to that from the $\tau \to 3
\mu$ decay.

We calculated the $K$-$\bar K$ mixing mass difference and the
$CP$-violating parameter $\epsilon_K$.  Due to the enhancement in the
left-right mixing terms, their coefficients are two orders of
magnitude bigger than those of purely left-handed and right-handed
terms.  Therefore, the constraint on the product $B_{ds}^L B_{ds}^R$
is much stronger than those on $(B_{ds}^L)^2$ and $(B_{ds}^R)^2$.  The
mass difference provides a limit on the real part of $B_{ds}^L
B_{ds}^R$, while the $\epsilon_K$ provides a limit on its imaginary
part.

We also evaluated the $B_d$-$\bar B_d$ mixing in the context of the
flavor-changing $Z'$ couplings.  Because the measured mass difference
and $CP$ asymmetry may be partially involve new physics at present, we
can no longer use the $V_{td}$ determined from the $B_d$-$\bar B_d$
system assuming only the SM physics.  Instead, $V_{td}$ is relexed to
all possible values allowed by the unitary triangle, with $|V_{ub}|$,
$|V_{cb}|$ and $|V_{cd}|$ fixed by the semileptonic $B$ decays and the
neutrino and atineutrino production of charm.  Furthermore, because
$Z'$ contributions to $b \to s \gamma$ and $b \to d \gamma$ decays are
both loop and mass suppressed, these processes can be used to
constrain the SM $|V_{td}/V_{ts}|$.  We used such limits to improve
the analysis on $B_d$ mixing.  We found that there is only a small
window for the $Z'$ physics in the $B_d$ system when one takes into
account all the constraints on $\Delta M_B$, $\sin2\beta$, $|V_{ub}|$,
and $|V_{td}/V_{ts}|$ given by different experiments.

\begin{acknowledgments}
  
  We thank Paul Langacker for helpful discussions.  C.-W. C. is
  grateful for the hospitality of the National Center for Theoretical
  Sciences at Hsinchu, Taiwan.  This research was supported in part by
  the U.S.~Department of Energy under Grants No DE-FG02-96ER40969 and
  in part by the National Science Council of Taiwan, R.O.C.\ under
  Grant No.\ NSC 94-2112-M-008-023-.

\end{acknowledgments}

\end{document}